# Magnetic Anomaly in Superconducting FeSe

D. Mendoza*, J. L. Benítez, F. Morales, and R. Escudero

Instituto de Investigaciones en Materiales, Universidad Nacional Autónoma de México. Ciudad Universitaria, Coyoacán Distrito Federal 04510. México.

*Corresponding autor: doroteo@servidor.unam.mx

Fax: 52 55 56 16 12 51

## Abstract

Synthesis, electrical and magnetic characterization of superconducting $FeSe_{0.85}$ compound is reported. An anomaly in the magnetization against temperature around 90K is observed. Magnetic characterization of a commercial compound with nominal FeSe stoichiometry is also presented. The overall magnetic behaviors as well as the magnetic anomaly in both compounds are discussed in terms of magnetic impurities and secondary phases.

## 1. Introduction

The discovery of superconductivity in iron selenide with tetragonal structure has generated a great deal of interest because it is a simple binary compound containing a ferromagnetic element [1]. The critical temperature is around 8 K, but can be increased with pressure reaching a maximum of about 37 K at 7 GPa [2-11]. It is found that in cooling, FeSe presents a structural transition from tetragonal to orthorhombic. The structural temperature transition was reported occurring at 105 K [1], 100 K [12], 90 K [13], and 70 K [14]. One important issue addressed since the discovery of this superconductor, is related to its stoichiometry. At the very beginning, superconductivity was reported occurring in the tetragonal phase, as $FeSe_x$ with a minimum value of x=0.82 [1], however, actually it has been stated that the superconducting compound is close to the ideal stoichiometry (x=1) [12, 14-16]. When the synthesis is performed with selenium deficiency, impurity phases appear, mainly hexagonal FeSe, elemental iron, and in some cases iron oxides.

An interesting aspect about this compound is related to the magnetic features above the transition temperature observed in the first paper [1], as well in several reports (see for example: [17-20]). The magnetic feature may be characterized as a bump at around 90 K, but was also observed as a step around 120 K-130 K [4, 7, 21]. Blachowski, et al. [20] observed both kinds of anomalies in their samples; a bump located from 80 K to 100 K, and a step around 125 K. They associated the former anomaly to a structural transition, and the last one to an antiferromagnetic type transition due to the presence of a hexagonal phase impurity; specifically to $Fe_7Se_8$ as indicated by their Mössbauer experimental analysis.

In this work we report the synthesis of the superconducting compound $FeSe_x$ with nominal selenium deficiency, x=0.85. At this stoichiometry the magnetic feature, the named bump, is observed; we will discuss the possible origin of this anomaly. For comparative purposes we also present the magnetic characterization of a commercial compound with nominal FeSe stoichiometry.

## 2. Experimental details

Two FeSe samples were used in this work. The first ones was prepared starting from iron and selenium (ESPI, 5N purity) powders, they were mixed and sealed in evacuated fused quartz ampoules to form the $FeSe_x$ compound with nominal concentration of x=0.85. The iron powder was obtained by reduction of hydrated iron nitrate (Tecsiquim, México, purity >98%) in a hydrogen atmosphere at $850^{O}C$. After 12 hr at $670^{O}C$, the $FeSe_{0.85}$ sample was cooled to room temperature. The mixture was grounded and pelletized, sealed again and set at $670^{O}C$ during 24 hr and then at $400^{O}C$ for 48 hr, and finally cooled to room temperature. For the second sample we used commercial (Alfa Aesar, 99.9% purity) compound with nominal concentration FeSe, this sample was characterized as received.

X ray diffraction analysis was performed using a Phillips D5000 diffractometer with CuKα line. Magnetization and resistivity measurements were made in a Quantum Design magnetometer MPMS5 and PPMS, respectively.

## 3. Results

Figure 1 shows the X ray diffraction patterns of the studied samples. The upper one is the XRD powder spectrum of the commercial FeSe sample, whereas in the lower one displayed is the synthesized $FeSe_{0.85}$. For $FeSe_{0.85}$, the majority phase corresponds to the tetragonal structure and the minority impurities can be identified as iron oxide ($Fe_2O_3$), hexagonal FeSe, and elemental iron. For the commercial FeSe sample the majority phase displayed is the tetragonal structure, as a secondary phase the hexagonal structures is observed, and the existence of elemental selenium is also clear. Note that the hexagonal phase is more evident in the commercial FeSe sample.

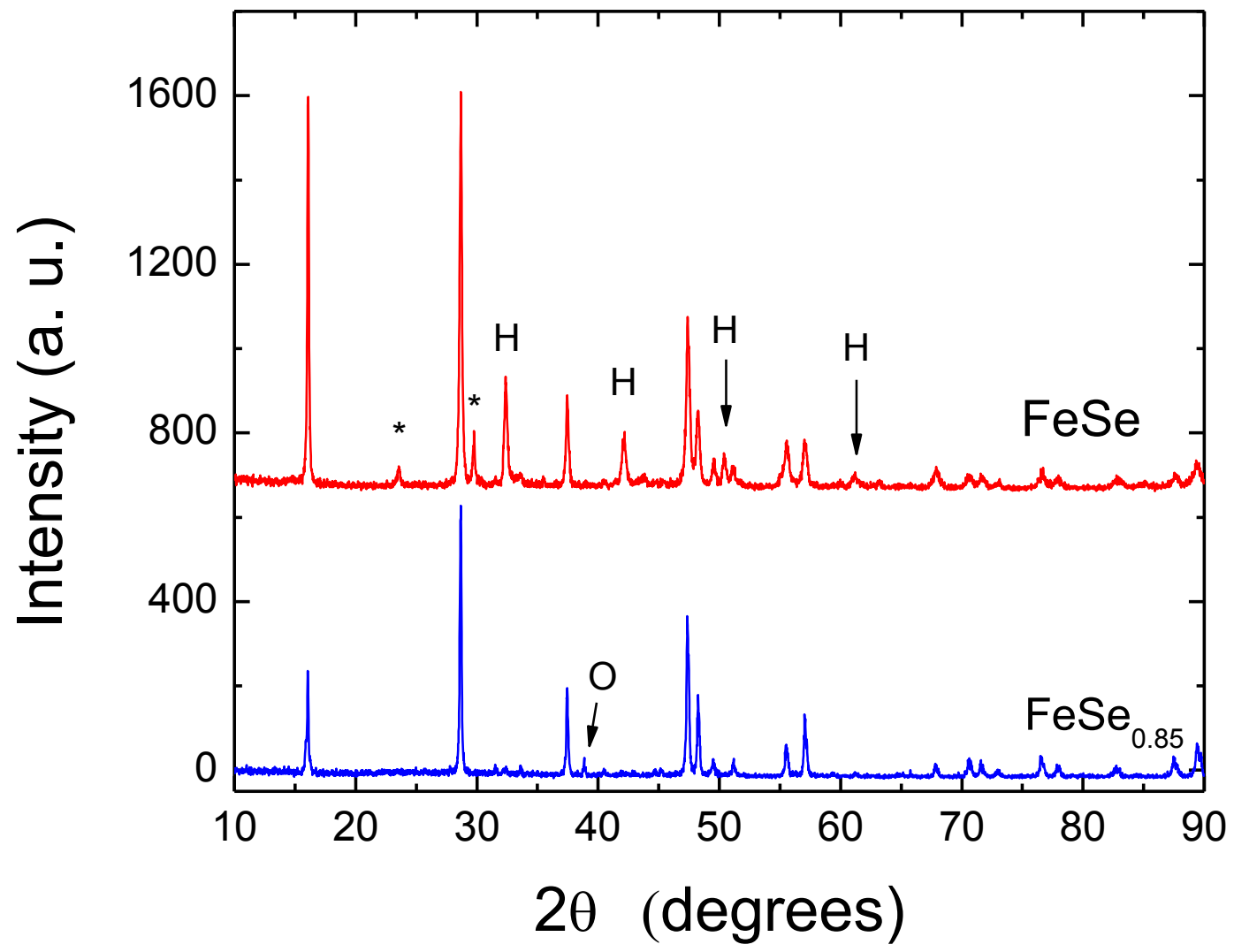

Figure 1. X-ray patterns of commercial FeSe and synthesized $FeSe_{0.85}$ samples. H is the symbol for hexagonal phase of FeSe, indexed to the $Fe_7Se_8$ structure; asterisks and O indicate elemental selenium and iron oxide, respectively.

Figure 2 shows the resistivity versus temperature, $R(T)$, for the $FeSe_{0.85}$ sample. There, two interesting characteristics are present; a superconducting transition temperature $T_C$ attained at 7.2 K (zero resistivity), and a change of slope is observed around 80 K.

In the two panels of Fig. 3 we present magnetization versus temperature, $M(T)$, for the two samples. Both panels show shielding and Meissner diamagnetic characteristics, as measured by Zero Field Cooling (ZFC), and Field Cooling (FC) cycles, under a magnetic field of 10 Oe. Above $T_C$ interesting features are displayed in both samples; a considerable paramagnetic signals and irreversibility anomalies. The upper panel ($FeSe_{0.85}$ sample) presents irreversibility in the ZFC and FC measurements at about 110 K, when decreasing temperature, the characteristic bump is clearly marked at 90 K. In the lower panel (commercial FeSe sample) a step is clearly observed at about 125 K in both ZFC and FC cycles.

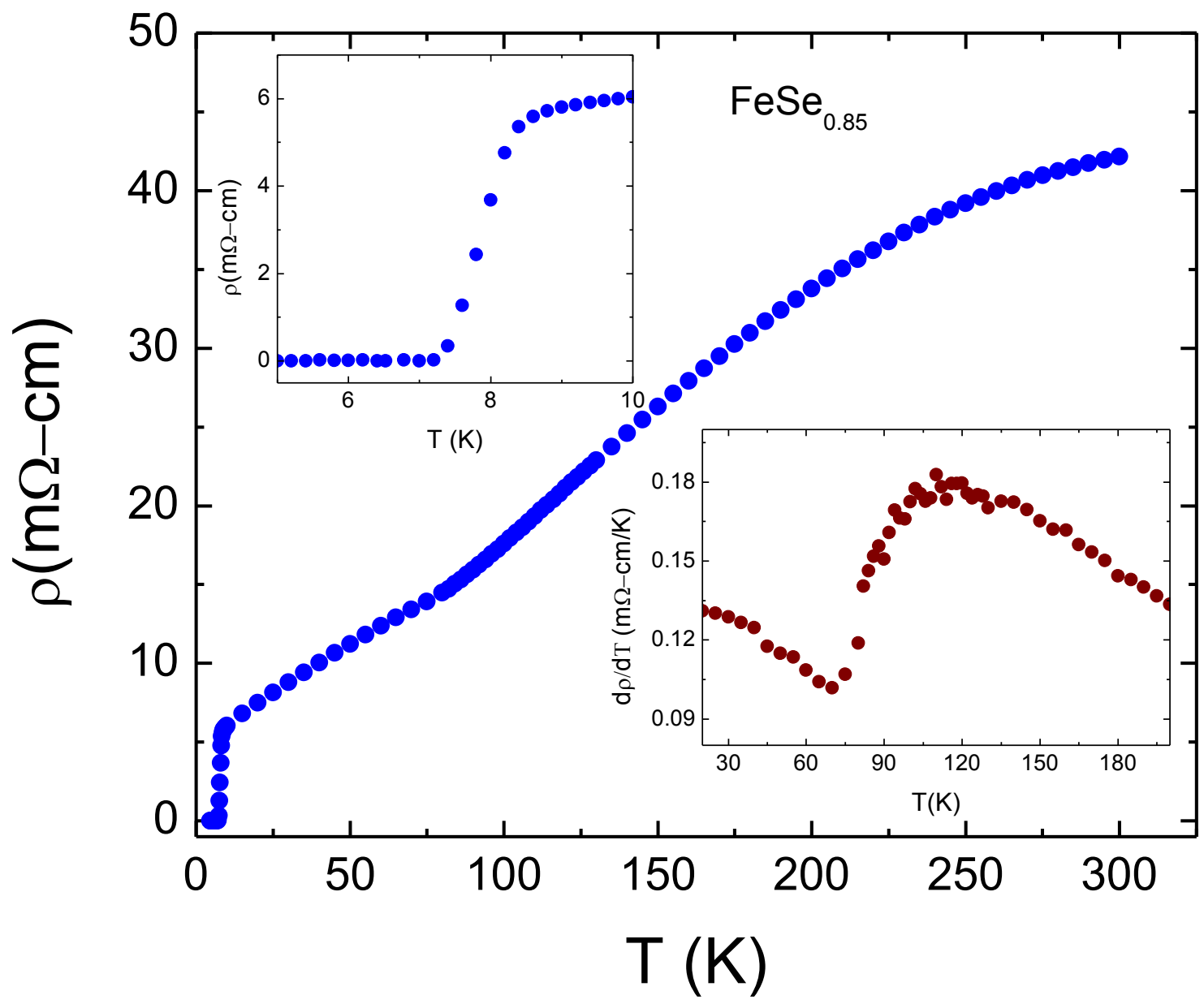

Figure 2. Resistivity against temperature of the $FeSe_{0.85}$. Upper inset shows the ρ(*T*) behavior at low temperatures where it reaches zero resistance around 7.2 K. Lower inset dρ/dT presents the change of slope around 80 K.

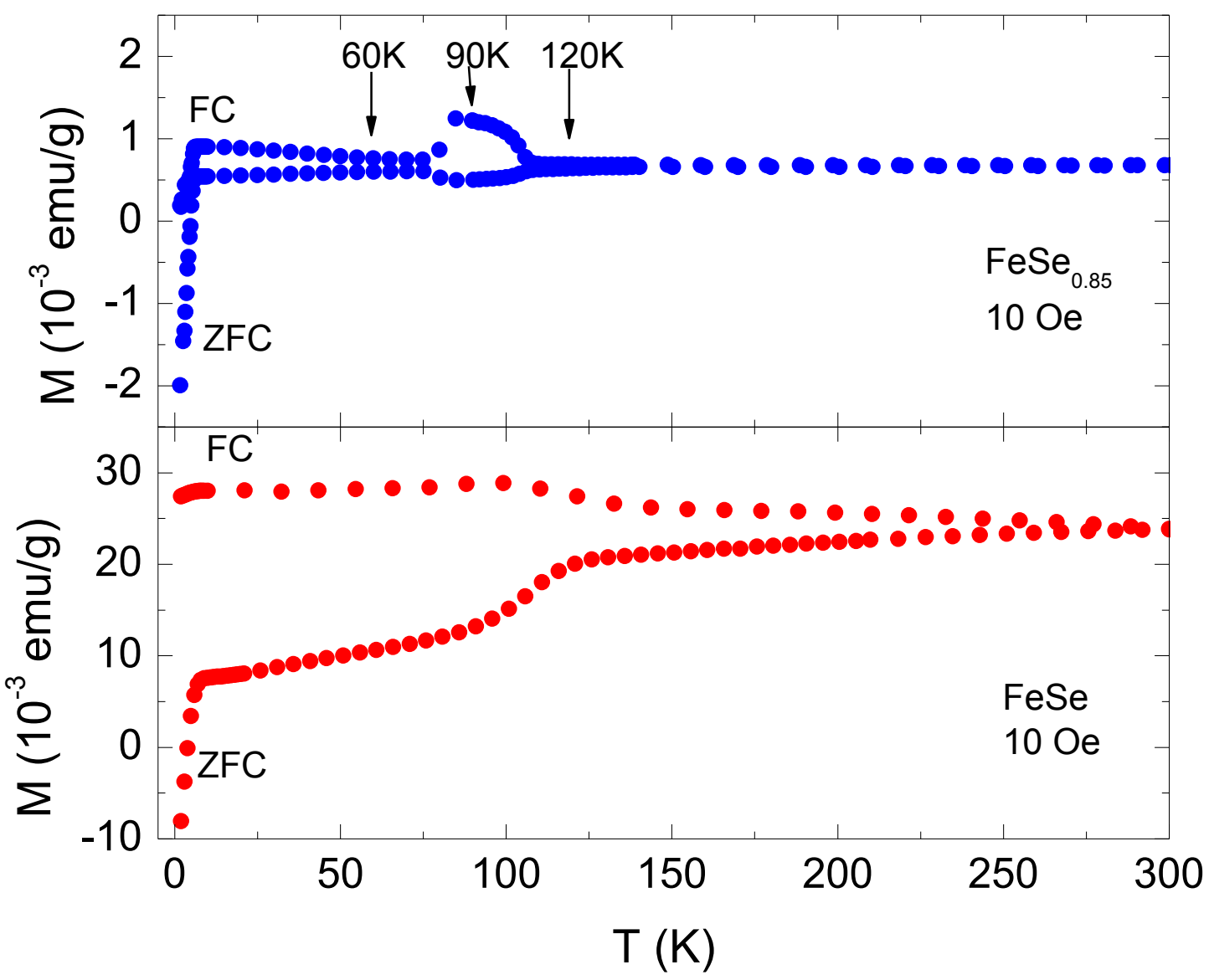

Figure 3. *M*(*T*) curves for $FeSe_{0.85}$ and commercial FeSe. The arrows indicate the temperature where isothermal *M*(*H*) measurements were performed.

In order to take some insight about the nature of the anomalies observed in the 75 K-110 K range in the $FeSe_{0.85}$ sample, we performed isothermal measurements of magnetization as a function of applied magnetic field *M*(*H*) at ±1 Tesla, at the three temperatures marked by the arrows in Fig. 3. Figure 4 displays *M*(*H*) measurements taken at 120 K, 90 K and 60 K. The overall form of these curves suggests a ferromagnetic behavior with very small hysteresis within the resolution of the instrument. The three curves panels 4a, 4b, and 4c collapse into the same curve as is shown in panel 4d. At first look, this might indicate that there is no magnetic transition through the *M*(*T*) anomaly, at least no detected by *M*(*H*) measurements.

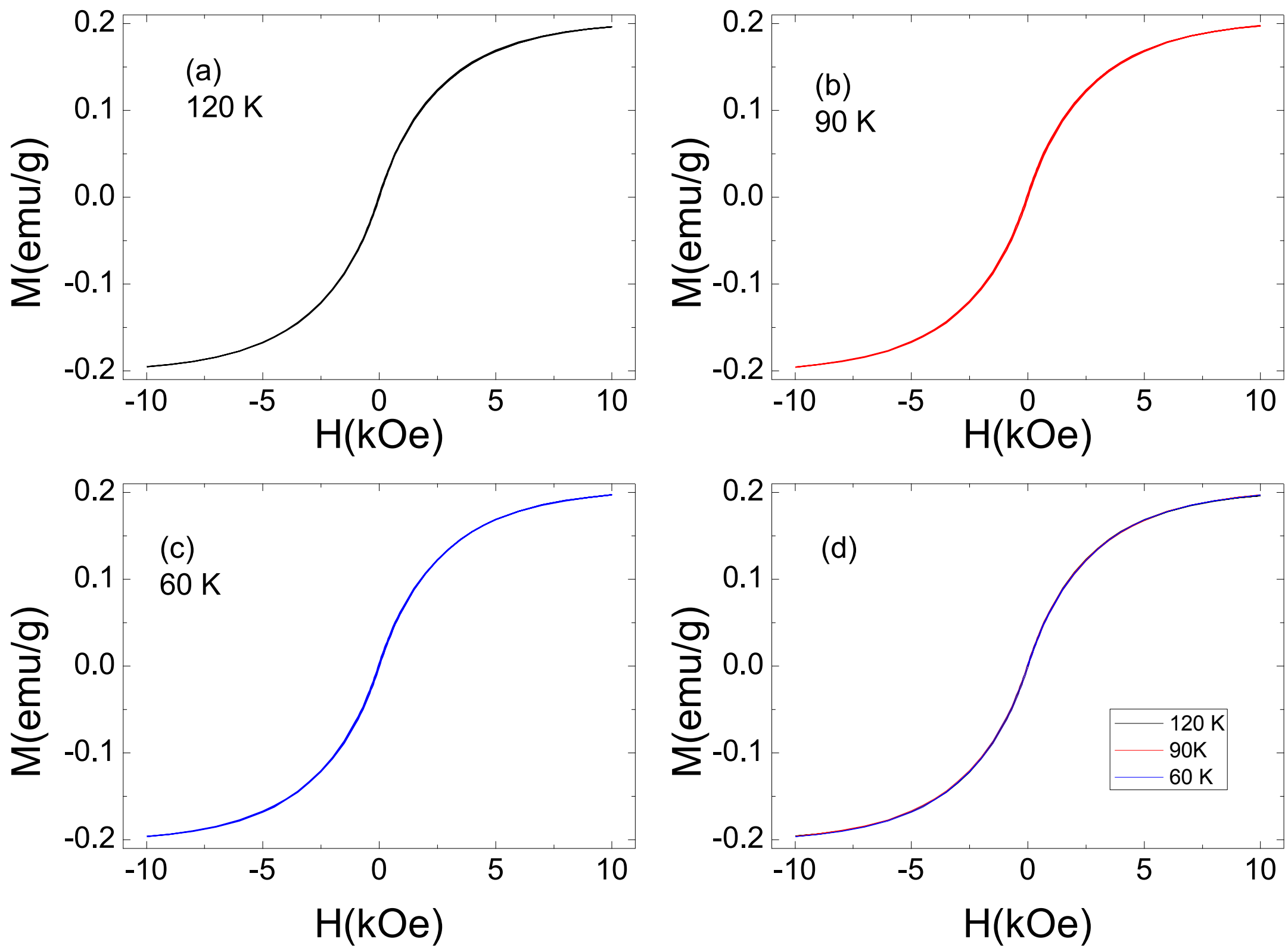

Figure 4. *M*(*H*) curves measured at different temperatures around the magnetic anomaly for the $FeSe_{0.85}$ sample: (a) 120 K, (b) 90 K, and (c) 60 K. In (d) all curves plotted in the same graph.

4. Discussion

In the majority of the reports concerning to the superconductivity in the FeSe compound, where some kind of anomaly in magnetization measurement is observed, the synthesized compounds are deficient in selenium and impurities phases are commonly reported. The question is whether the impurity phases induce those anomalies and the positive background or the tetragonal FeSe is intrinsically magnetic.

Theoretical calculations show that tetragonal FeSe is not magnetic but magnetism is present in its hexagonal phase [22]. Previous experimental measurements indicate magnetic behavior in tetragonal FeSe, with Curie temperature around 325 K, but the magnetism is attributed to Fe clusters and Fe vacancies in Fe rich and Se rich compounds, respectively [22, 23]. More recent calculations on tetragonal selenium deficient $FeSe_x$ compound show that magnetism is located around Se vacancies [24]. On the other side, hexagonal FeSe has been shown to be ferrimagnetic with Néel temperature around 420 K [25]. A related hexagonal $Fe_7Se_8$ compound shows ferrimagnetism below 450 K and antiferromagnetic-like transition around 130 K [26]. In a recent report [16], studies on the $FeSe_x$ compound with $0.82 \leq x \leq 1.14$ indicates the existence of impurity phases that destroys the superconductivity (compared to the ideal case x=1), and the appearance of magnetism. The limit case x=1.14, corresponding to the stoichiometry $Fe_7Se_8$, shows the strongest magnetic signal.

Superconducting FeSe has resulted to be very sensitive to the stoichiometry [15], and even in a most cleaned sample ($FeSe_{0.99}$) a clear magnetic behavior is observed [16]. Hence, experimental and theoretical evidence indicate that impurity phases and/or defects (vacancies, interstitials) might produce some kind of magnetism in the tetragonal FeSe compound.

Returning to the magnetic anomaly observed around 90 K in the superconducting compounds, Mössbauer spectroscopy has been used to study this material [13, 20, 27], and no conclusive evidence of some kind of magnetic transition have been observed around such anomaly. This is consistent with measurements presented in figure 4, where no appreciable change in the $M(H)$ curves is notable through the temperature range where the magnetic anomaly is observed. Our $FeSe_{0.85}$ sample contains traces of magnetic compounds (iron oxide, Fe, hexagonal FeSe). We believe that the anomaly is produced by magnetic impurities, due to an environmental change that surrounds those impurities by the tetragonal-orthorhombic structural transition. The temperature of the structural transition by itself may be affected by the kind and location of the impurity into the crystalline structure, as is appreciated by the wide interval of temperatures (70 K-105 K) where the transition is observed [1, 12-14]. In our case, the magnetic anomaly is located within 75 K-110 K interval (see figure 3), and a change in the slope of resistivity against temperature is observed at about 80 K (see lower inset in figure 2), which may also related to the structural transition.

We think that the positive background observed in the whole $M(T)$ characteristics is also produced by magnetic impurities (of course, besides the usual Pauli contribution). This can be corroborated in the $M(H)$ curves presented in figure 4, where a typical ferromagnetic-like behavior is observed, although very small hysteresis is appreciated. It should be noted that ferromagnetic-like behavior have been reported in superconducting FeSe by means of $M(H)$ measurements at temperatures above $T_C$ [16, 28, 29], below $T_C$ [29] and even in a very pure sample [16].

Other kind of anomaly commonly observed in the superconducting FeSe samples is a step-like change in the magnetization curve in the ZFC measurement ending around 120 K-130 K [4, 7, 21], similar to that observed in figure 3 for the commercial sample. Blachowski, et al. [20] have reported this kind of anomaly and by means of Mössbauer experiments analysis they adjudicated it to the presence of a hexagonal FeSe phase, specifically to spin rotation transition around 125 K in the hexagonal $Fe_7Se_8$ compound.

It has been reported that selenium excess induces the formation of the hexagonal $Fe_7Se_8$ structure [16]. Figure 1 shows the existence of elemental selenium for the commercial FeSe, which may indicate that this compound was synthesized with an excess of Se respect to elemental iron. XRD analysis (see figure 1) shows that the commercial FeSe sample contains considerable quantities of a hexagonal phase, which can be indexed to the $Fe_7Se_8$ stoichiometry (PDF card 00-048-1451). The step observed in $M(T)$ of this sample (see figure 3), as Blachowski, et al. [20] propose, may be related to the presence of this kind of magnetic phase. It is interesting to note that both $FeSe_{0.85}$ and commercial FeSe compounds are magnetic at room temperature (as was readily tested by using a small permanent magnet). This fact may be explained by the presence of both Fe and hexagonal FeSe impurities in the $FeSe_{0.85}$ sample, and by the hexagonal $Fe_7Se_8$ phase in the commercial sample. Recall that experiments show that tetragonal FeSe with impurities [23], hexagonal FeSe [25], and $Fe_7Se_8$ [26] are magnetic at room temperature.

Finally, the analysis presented here is for polycrystalline FeSe samples. To have a complete picture on the magnetic behavior it is necessary to study pure single tetragonal FeSe samples. Some attempts have been made to this respect (see for example: [8, 29-33]), but synthesized crystals contains considerable traces of a hexagonal phase. In one report, magnetization measurements with external magnetic field parallel to the c axis and parallel to the a-b plane have been made, showing small irreversibility in the FC and the ZFC processes [33]. In this case, a broad maximum located between 150 K-200 K is observed when the external magnetic field is parallel to the a-b plane. Similar broad maximum in the $M(T)$ curves have been reported by others [8, 30], but the orientation of the sample in the measurement is not specified. Zhang, et al. [31] report a very broad maximum prolonged to higher temperatures with no appreciable irreversibility measured with the field parallel to the c axis. We believe that, at least, the observed broad maximum might be explained if one assumes that the hexagonal phase detected in those samples is associated to the $Fe_7Se_8$ structure. As mentioned before, the hexagonal $Fe_7Se_8$ phase presents an antiferromagnetic transition, around 120 K-130 K depending on the impurities [26]. Hence, for the reported magnetization measurements in crystals, we believe that the overall form and the specific location of the transition would depend on the quality of the crystal and the interaction with the majority tetragonal FeSe phase that surrounds the hexagonal phase, along with the orientation of the sample respect to the magnetic field. But more work is necessary to this respect.

5. Conclusion

Superconducting $FeSe_{0.85}$ sample has small traces of magnetic impurities (iron oxide, Fe and hexagonal FeSe). Commercial FeSe is mainly in the tetragonal structure containing hexagonal FeSe that can be indexed to the $Fe_7Se_8$ compound as a secondary phase. The overall magnetic behavior of both samples looks different. Presenting a magnetic anomaly around 90 K, in the former, whereas in the last one, a step at around 120 K. Both samples have a large paramagnetic background in $M(T)$ characteristics. In $FeSe_{0.85}$ and by means of $M(H)$ measurements we did not find any significant difference in those characteristics through the anomaly, which might indicate that there is not an intrinsic magnetic transition. We propose that the anomaly in both samples is induced by the presence of magnetic impurities. We believe that the anomaly in the $FeSe_{0.85}$ is produced by a kind of magnetic fluctuation induced by the structural transition. In the case of commercial FeSe the magnetic behavior may be produced by a major presence of the magnetic hexagonal secondary phase.

Acknowledgements.

We wish to thank F. Silvar for technical support.

References.